\documentclass[apl, aps4, twocolumn, groupedaddress]{revtex4}
\usepackage{amsmath}
\usepackage{amsfonts}
\usepackage{amssymb}
\usepackage{graphics}
\usepackage{graphicx}
\usepackage{color}
\usepackage[applemac]{inputenc}
\usepackage[cyr]{aeguill}
\usepackage{psfrag}

\newcommand{\ii}{\text{i}}

\definecolor{green2}{rgb}{0,0.66,0} 

\begin{document}

\title{Broadband Coherent Perfect Absorption of Acoustic Waves with Bubble Meta-Screens}
\author{Maxime Lanoy}
\author{Reine-Marie Guillermic}
\author{Anatoliy Strybulevych}
\author{John H. Page}

\affiliation{Department of Physics and Astronomy, University of Manitoba, Winnipeg, Manitoba R3T 2N2, Canada}


\date{\today}

\begin{abstract}
A bubble meta-screen is an exceptionally effective and low frequency resonator which can be optimized in order to exactly balance the energy provided by radiative process and lost under a viscous mechanism (critical coupling). Under this condition, one can absorb 99.9\% of the energy carried by two phase-matched counter propagating acoustic beams. This phenomenon, called coherent perfect absorption, is here observed with bubbles 75 times smaller than the incident wavelength and is shown to be remarkably broadband. Finally, tuning the relative phases of the two beams turns out to be an efficient way to control the absorption in the medium.
\end{abstract}


\maketitle

The propagation of a wave is ruled by causality. As a direct consequence, a finite sized support is required to absorb its energy~\cite{yang2017optimal}.
Because subwavelength resonators provide a solution to confine the field in space regardless of the incident wavelength, they offer a way to overcome the classical causality related limitations and open up the way to the \textit{super-absorption} phenomenon. 
For instance, it has been shown that, for a unidimensional configuration, a single open resonator could by itself absorb up to 50\% of the incoming energy~\cite{bliokh2008colloquium} under the so-called critical coupling condition. The recent developments in the field of metamaterials offer new perspectives in this regard. Indeed, the high degree of control and the tunability of such structures enable a very flexible design of their resonant properties and yield useful applications such as negative refraction~\cite{shelby2001experimental,valentine2008three,lanoy2017acoustic}, invisibility cloaks~\cite{schurig2006metamaterial,chen2007acoustic,torrent2008acoustic,cummer2008scattering} or super-focusing~\cite{lerosey2007focusing,lemoult2011acoustic,lanoy2015subwavelength}. Also, some very convincing demonstrations of super-absorption have recently been achieved in optics~\cite{aydin2011broadband} as well as in acoustics~\cite{yang2010acoustic,mei2012dark,leroy2015superabsorption,romero2016perfect}. 
For example, a design consisting of a meta-screen of air bubbles trapped in a visco-elastic matrix~\cite{leroy2015superabsorption} can dissipate up to half the incident energy in transmission over a significantly broad range of frequencies despite the bubble radius being approximately a hundredth of the incident wavelength. Better yet, the absorption process can be fully completed by breaking the symmetry of the configuration and adding an opaque barrier at one end of the resonator.\\
The phenomenon of coherent perfect absorption (or anti-laser effect), first evidenced in optics~\cite{chong2010coherent} and more recently considered for acoustic waves~\cite{song2014acoustic,wei2014symmetrical, achilleos2016coherent, meng2017acoustic}, brings a new ingredient to the discussion. Indeed, the efficiency of this technique, which consists in fully absorbing the energy carried by two oppositely directed beams, has been shown to rely critically on the symmetry of the system and more specifically on the phase shift between the two incident beams.
\\

In this letter, we experimentally demonstrate the possibility of achieving coherent absorption of sound waves using a bubble meta-screen. Thanks to a previously developed model~\cite{leroy2009transmission}, we show that a meticulous design of the structure can lead to an optimized absorption over a broad range of frequencies. For this to happen, the viscous dissipation needs to exactly balance the radiative damping (critical coupling condition). After fabrication, we report absorption performances as high as 99.9\% at 1.8~MHz. More interestingly, broadband absorption greater than 98\% is demonstrated between 1.4 and 3.2~MHz. Finally, we show that the mechanism highly depends on the input symmetry. We can smoothly tune the absorption between 24\% and 99.9\% by switching from anti-symmetrical to symmetrical excitation.\\

\begin{figure}[htb!]
    \centering
      \includegraphics[width=.85\linewidth]{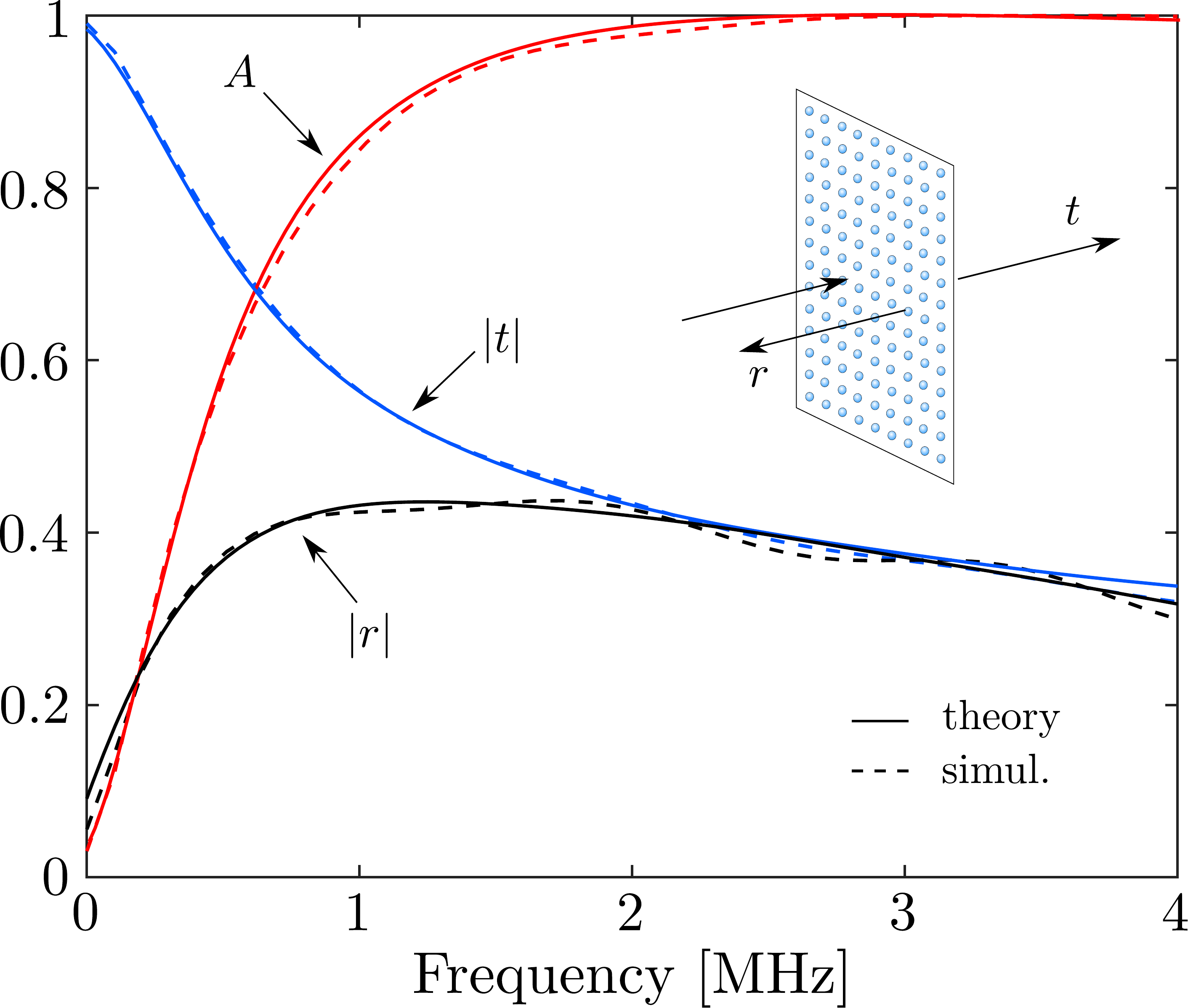}
    \caption{Theoretical (solid lines) and numerical (dashed lines) predictions for the transmission, reflection and absorption induced by a bubble metascreen with an optimized set of parameters (bubble radius $a=11~\mu$m, lattice parameter $d=d^*=58.1~\mu$m, shear viscosity $\eta=0.9$ Pa.s and bulk modulis $\mu=6$~MPa of the elastic matrix). The bubble layer is placed in the middle of the metascreen.}\label{opt}
\end{figure}

\begin{figure*}[htb!]
    \centering
      \includegraphics[width=.9\linewidth]{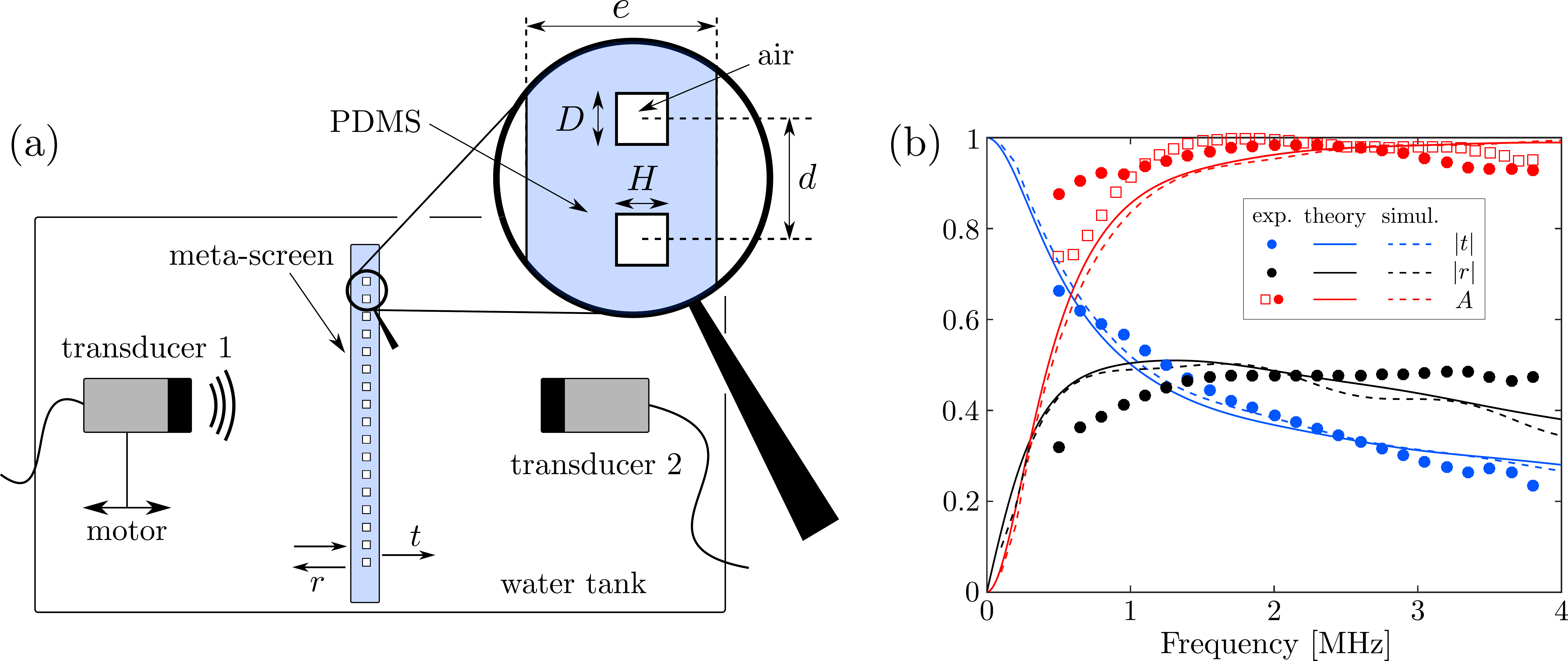}
    \caption{(a) Scheme of the experiment. A meta-screen consisting in a layer of cylindrical air cavities (diameter $D=24~\mu$m, height $H=13~\mu$m, equivalent radius $a=11~\mu$m) arranged within a square lattice (parameter $d=50~\mu$m) and trapped in a silicone (PDMS) matrix is immersed between a pair of identical transducers, one of which can be translated thanks to a motorized stage. (b) Experimental (symbols), theoretical (solid lines) and numerical (dashed lines) predictions for the transmission, reflection and absorption induced by our bubble metascreen. The reflection coefficient $r$ (black empty circles) and the absorption $A$ (red solid
circles) are extracted by exploiting the pressure continuity. The second experimental data set for $A$ (empty squares) corresponds to measurement performed under the real CPA conditions (both transducers emitting simultaneously) exploiting formula~(\ref{abs}).}\label{setup}
\end{figure*}
When undergoing an acoustic excitation, a bubble oscillates. For low frequencies, the oscillations remain essentially monopolar and exhibit resonant behaviour most commonly known as the Minnaert resonance~\cite{minnaert}. In water, the Minnaert resonance occurs for an incident wavelength 300 times larger than the size of the bubble thus making it a very promising candidate for the realization of thin locally resonant meta-materials. Our structure (see inset of fig.~\ref{opt}) consists of a layer of air bubbles of radius $a$, arranged on a square lattice of lattice constant $d$ and trapped in a polydimethylsiloxane (PDMS) viscoelastic silicone rubber slab with shear viscosity $\eta=0.9$ Pa.s and modulus $\mu=6$~MPa. To avoid spurious reflections at the water-PDMS interface, impedance matching was ensured by loading the polymer with a 17.5\% volume fraction of TiO$_2$ nanoparticles as explained in~\cite{guillermic2018pdms}. The particles are found to increase the longitudinal attenuation in th bulk PDMS taking the form $\alpha_\text{m}=0.2 f^{1.38}$ mm$^{-1}$ ($f$ in MHz). The total thickness of our meta-structure being $e=630~\mu$m, 24\% of the overall energy is expected to be dissipated by the TiO$_2$-PDMS matrix at 1.8~MHz.\\

When several bubbles coexist, they couple via the phenomenon of multiple scattering. As a consequence, the description of a bubbly medium is a lot more complex than that of a single bubble. Yet, in our particularly simple geometry, it has been shown~\cite{leroy2009transmission} that one can successfully analytically model the reflection and transmission coefficients $r_\text{bub}$ and $t_\text{bub}$  describing the interaction of the bubbly layer with an incident plane wave of angular frequency $\omega$ and wavenumber in water $k$. More strikingly, it occurs that the bubble screen can itself be described as an open resonator. Introducing $K=2\pi/(k d^2)$, which represents the ability of the layer to radiate in the longitudinal direction, and $I=1-Ka\text{sin}(k d/\pi)$, which accounts for the coupling between neighboring cavities, one can show that the bubbly layer resonance occurs at $\omega_0=\omega_M/\sqrt{I}$. Finally, we have the following expressions:
\begin{align}
r_\text{bub} & = \frac{\ii  Q_\text{rad}^{-1} x }{1- x^2-\ii  Q^{-1} x} \\
\text{and} \hspace{8pt} t_\text{bub} & =  1+ \frac{\ii  Q_\text{rad}^{-1} x }{1- x^2-\ii  Q^{-1} x} ,
\end{align}
where $x=\omega/\omega_0$, $Q^{-1} =  Q_\text{rad}^{-1}+ Q_\text{diss}^{-1}$ stands for the inverse quality factor of the bubble layer and accounts for both the radiative damping $Q_\text{rad}^{-1}=Kax/I$ and the dissipation $Q_\text{diss}^{-1}=4\eta/(\rho a^2 \omega_0 I)$ which, in this case, essentially originates from viscous mechanism.
The actual reflection and transmission through our metascreen can be derived from the above expressions by taking into account the effect of the longitudinal attenuation in the matrix: $r=r_\text{bub}   \exp(-\alpha_\text{m} e/2)$ and $t=t_\text{bub}   \exp(-\alpha_\text{m} e/2)$.
In a coherent perfect absorption configuration, the sample is excited from both sides. The output signal thus results from the interference between the transmitted and reflected channels.
One can then estimate the total absorption (for symmetrical inputs) as follows~\cite{wei2014symmetrical}:
\begin{equation}
A=1-|r+t|^2.
\label{abs0}
\end{equation}
At the meta-screen resonance, \textit{i.e.} for $x=1$, the absorption is $A =1- | Q_\text{dis}^{-1}- Q_\text{rad}^{-1}|^2 \exp(-\alpha e)/| Q^{-1}|^2$ meaning that perfect absorption is obtained for $Q_\text{dis}= Q_\text{rad}$. This phenomenon, commonly referred to as critical coupling~\cite{xu2000scattering}, occurs as the viscous damping exactly balances the radiative leaks of the meta-screen. In the present case, the criterion is satisfied for an optimal lattice parameter $d^*=(\pi Z a^3/2 \eta)^{1/2}$, where $Z$ is the acoustic impedance of the matrix. Because $d^*$ does not explicitly depend on frequency, the optimization is expected to be robust over a wide range of frequencies. Considering the characteristics of our impedance matched PDMS and setting the bubble radius to $a=11~\mu$m, one finds an optimal lattice parameter of $d^*=58.1~\mu$m. Fig.~\ref{opt} reports theoretical and numerical (FEM simulations) predictions for the performances for such an optimized metascreen. Besides the excellent agreement between the model and the simulations, one observes that an overall absorption of 100~\% can be expected at 3~MHz, and that the phenomenon is very broadband ($A>99$\% for $2.2~\text{MHz}<f<4.4~\text{MHz}$ for these parameters).\\

In this letter, we propose to explore experimentally the performance of a nearly optimized design. Thanks to soft photolithographic techniques, we were able to realize cylindrical air cavities of diameter $D=24~\mu$m and height $H=13~\mu$m in the impedance-matched PDMS. In the regime we are investigating, the incident wavelength is several orders of magnitudes greater than the size of such a cavity. As a consequence, they are expected to behave like spherical bubbles of equivalent volume, \textit{i.e.} with a  radius $a=(3D^2H/16)^{1/3}=11~\mu$m~\cite{meyer1958pulsation,leroy2009design,calvo2012low}. One can thus estimate the Minnaert resonance of a single cavity from the formula $\omega_\text{M}=\sqrt{3\beta_g+4\mu/(\rho a^2)}$~\cite{alekseev1999gas}; where $\beta_g$ stands for the bulk modulus of the gas and $\rho$ is the density of the outer matrix. A bubble metascreen could be achieved by arranging such inclusions within a square lattice of parameter $d=50~\mu$m (slightly smaller than our prediction for the optimal separation $d^*$). Despite this small discrepancy, a simple transmission experiment, realized by adopting the configuration of Fig.~\ref{setup}(a) where transducer 1 emits and transducer 2 receives, gives a very positive picture of the performance one can expect here. The results are shown in Fig.~\ref{setup}(b). The transmission coefficient (black solid circles), in agreement with the prediction from Eq.~(2), can be used to estimate the reflection coefficient (empty circles) by exploiting the pressure continuity at the center of the bubble layer. This allows us to circumvent the difficulty of realizing a reliable reference in reflexion; which is required to perform the direct measurement of $r$. One can then obtain the absorption (solid red circles) which turns out to be significant over the whole 1~MHz to 3~MHz bandwidth and features a maximum at 1.8~MHz.\\

\begin{figure}[htb!]
    \centering
      \includegraphics[width=.9\linewidth]{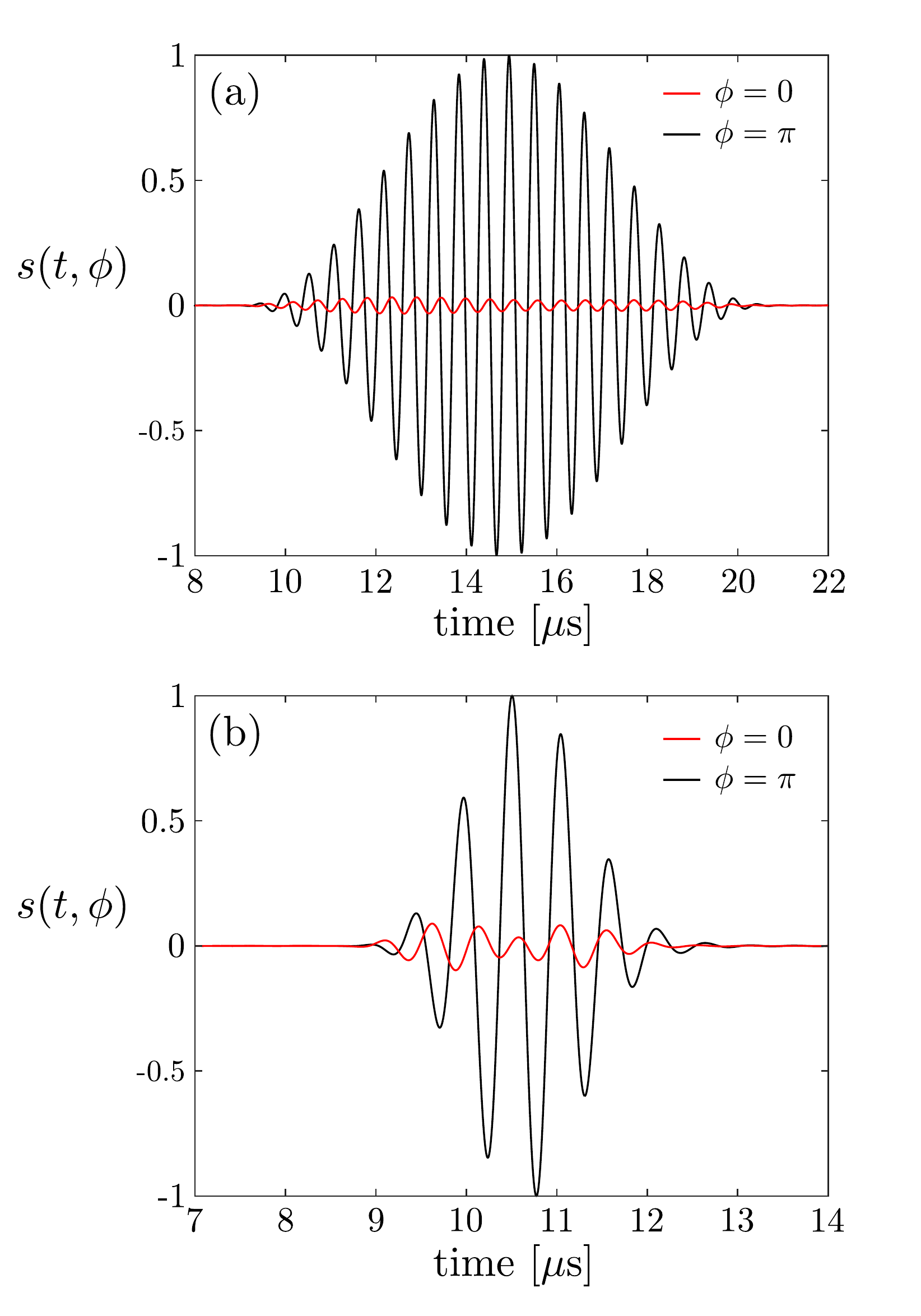}
    \caption{Time acquistion of the pressure fields collected at transducer 2 when both transducers are emitting the same narrowband (a) or broadband (b) pulse centered at 1.8~MHz. We explore both symmetrical ($\phi=0$ in red) and anti-symmetrical ($\phi=\pi$ in black) inputs configurations. Signals are normalized by the same amount.}\label{time}
\end{figure}
The actual coherent perfect absorption experiment is then performed by emitting from both transducers and receiving with transducer 2. The phase shift $\phi$ between both incident beams can be controled by mounting transducer 1 on a motorized translation stage. Our time measurements, as reported in Fig.~\ref{time}, provide a good picture of the absorption performance. We first examine the case of a long pulse centered on 1.8~MHz (see  Fig.~\ref{time}(a)). This kind of excitation has a relatively narrow frequency content meaning that we essentially measure $s_\omega(\phi)=s_0\big[r+t\exp(\ii\phi)\big]$ with $s_0$ the complex amplitude provided by the sources. For frequencies above $\omega_0$,  $r$ and $t$ are intrinsically opposed in phase meaning that an asymmetric input ($\phi=\pi$) will result in a constructive interference between the two channels. The total amplitude is then driven by the dissipation in the matrix: $s_\omega(\pi)=s_0 \exp(-\alpha_\text{m} e /2)$.  One can then generalize Eq.~(\ref{abs0}) to obtain a $\phi$-dependent expression for the quasi monochromatic absorption:
\begin{equation}
A_\omega(\phi)=1-\left| \frac{s_\omega(\phi)}{s_\omega(\pi)} \right|^2\exp(\alpha_\text{m} e).
\label{abs}
\end{equation}
For the situation of Fig.~\ref{time}(a), we report a maximum absorption as high as $A_\omega(0)=99.9$\%. As a matter of fact, as depicted on Fig.~\ref{time}(a), the signal almost vanishes when the input becomes symmetrical (red line).
We also demonstrate the robustness of our optimization with frequency by reporting the case of a significantly shorter pulse (see Fig.~\ref{time}(b)). Once again the amplitude reduction is considerable in the case of a symmetrical input ($\phi=0$). This is, to our knowledge, the first experimental demonstration of broadband coherent perfect absorption of acoustic waves so far.
This measurement also provides an alternative estimate for the absorption evolution with frequency. Note that, because of fabrication limitations, the bubbles could not be placed exactly at the center of the sample, which resulted in a slight drift of the phase shift $\phi$ when modifying the driving frequency. This effect is not detrimental to the broadband nature of the phenomenon. Yet, one needs to apply the appropriate phase correction to obtain the evolution reported on Fig.~\ref{setup}(b) (open red squares) which is consistent with the projection obtained from the transmission measurement (red circles) and is also well described by the trend predicted by the analytical model (red line).~~\\~~\\

\begin{figure}[htb!]
    \centering
      \includegraphics[width=\linewidth]{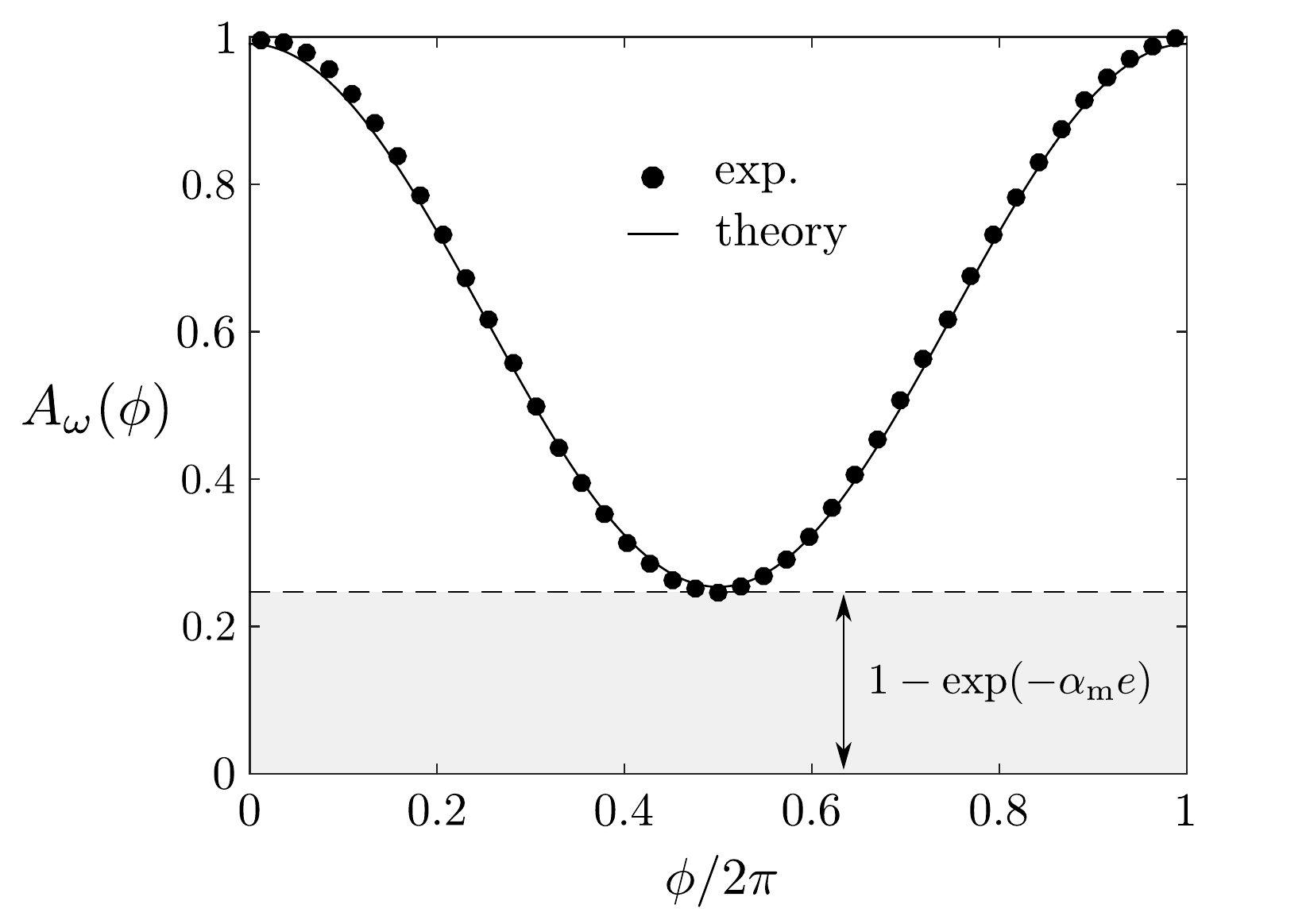}
    \caption{Coherent control. The phase shift between both incident beams was modified by translating transducer 1. For each position, we send a quasi-monochromatic excitation and estimate the absorption via Eq.~(\ref{abs}) (symbols). The gray shaded area pictures the inherent longitudinal attenuation of the medium.}\label{coherent}
\end{figure}
Finally, we provide an illustration for the coherent control of the absorption (see Fig~\ref{coherent}). By translating transducer 1, we can vary the phase shift between the two input beams and estimate the overall absorption $A_\omega(\phi)$ (symbols) via Eq.~(\ref{abs}). The waveform consists of a narrowband 1.8-MHz-centered pulse. Here, we were able to tune the overall absorption from 24\% to 99.9\%. We find the same behavior by injecting the analytical coefficients in the expression $1-||r|+|t|\exp(\ii\phi)|^2$ (solid line). Note that the lower limit corresponds to the attenuation in the bulk of the silicone matrix and is directly related to the sample thickness. By improving the fabrication process, one could realize thinner samples and significantly decrease this lower attenuation limit. The upper limit as well as the broadband aspect can also be further improved by picking a different elastomer or by modifying the lattice constant of the crystal.
~~\\

In conclusion, it was shown that bubbly meta-materials are excellent candidates for coherent perfect absorption of acoustic waves. The phenomenon, traditionally believed to be limited to narrowband inputs~\cite{longhi2010backward} is here found to be robust over a broad frequency range. Also, thanks to the low frequency nature of the Minnaert resonance, our meta-structure turns out to be very efficient despite being smaller than the incident wavelength. Finally we show that the phase shift between both incident beams provides a very efficient way to tune the absorption. Having demonstrated the possibility of using bubbles metascreens for superabsorption~\cite{leroy2015superabsorption} or rheometry~\cite{lanoy2018phononic} purposes, we propose that the phase dependence of A and the extreme sensitivity of CPA to the position of the meta-layer can be exploited to design interferometry based vibrometers and switches. In the system studied here, the sensitivity can be on the order of $1~\mu$m. Moreover, the broadband character makes it possible to envisage either continuous wave or pulse applications.
For example, since the critical coupling condition can also be achieved under oblique incidence~\cite{leroy2015superabsorption}, our device could be used as a switch for inputs with different incident directions and bandwidths.
~~\\

The authors would like to thank Eric Lee, Fabrice Lemoult and Valentin Leroy for their valuable inputs on simulation and theory. We acknowledge support from the NSERC Discovery Grant program.

%

\end{document}